%% file: neurips_2024.tex
\documentclass{article}

\PassOptionsToPackage{numbers, compress}{natbib}

\usepackage[preprint]{neurips_2024}

\usepackage[utf8]{inputenc} 
\usepackage[T1]{fontenc}    
\usepackage{url}            
\usepackage{booktabs}       
\usepackage{amsfonts}       
\usepackage{nicefrac}       
\usepackage{microtype}      
\usepackage{xcolor}         

\usepackage{amsmath}
\usepackage[pdftex]{graphicx}

\usepackage{changepage} 
\usepackage{threeparttable, tablefootnote}  
\usepackage{algorithm,algcompatible}  
\usepackage{comment} 
\usepackage{wrapfig} 

\newtheorem{proposition}{Proposition}
\newtheorem{definition}{Definition}

\title{Neural Polarization: Toward Electron Density for Molecules by Extending Equivariant Networks}

\author{
  Bumju Kwak \\
  Independent Researcher \\
  \texttt{meson3241@gmail.com} \\
  \And
  Jeonghee Jo* \\
  Korea Institute of Science and Technology (KIST) \\
  \texttt{jade.jeonghee.jo@gmail.com} \\
}

\begin{document}

\maketitle

\begin{abstract}
  Recent SO(3)-equivariant models embedded a molecule as a set of single atoms fixed in the three-dimensional space, which is analogous to a ball-and-stick view. This perspective provides a concise view of atom arrangements, however, the surrounding electron density cannot be represented and its polarization effects may be underestimated. To overcome this limitation, we propose \textit{Neural Polarization}, a novel method extending equivariant network by embedding each atom as a pair of fixed and moving points. Motivated by density functional theory, Neural Polarization represents molecules as a space-filling view which includes an electron density, in contrast with a ball-and-stick view. Neural Polarization can flexibly be applied to most type of existing equivariant models. We showed that Neural Polarization can improve prediction performances of existing models over a wide range of targets. Finally, we verified that our method can improve the expressiveness and equivariance in terms of mathematical aspects.
\end{abstract}

\input{introduction}
\input{preliminary}
\input{method_and_experiment}

\input{result_table}

\input{result}
\input{discussion}
\input{conclusion_and_limitation}


\bibliography{neurips_2024}


\newpage
\appendix

\section{Appendix / supplemental material}

\setcounter{figure}{0}                       
\renewcommand\thefigure{S\arabic{figure}}   
\setcounter{table}{0}                        
\renewcommand\thetable{S\arabic{table}}     


\input{related_work}
\input{dft}
\input{equivariance}

\input{other}

\end{document}

%% file: introduction.tex
\section{Introduction}

For chemical engineering, accurate estimation of molecular conformation with electronic configuration is an essential factor. Quantum chemistry \cite{levine2014quantum, bartok2010gaussian} is a branch of chemistry of studying quantum mechanics of a molecule conformation, based on microscopic analysis of a single atom and its surroundings. The most common approach of quantum mechanical modeling of a molecule is density functional theory (DFT) \cite{mardirossian2017thirty}. From this perspective, the main strategy for solving this equation is considering the many-electron system as a functional for a single function, which corresponds to an electron density of a molecule in three-dimensional space \cite{yang1991direct, kryachko2012energy}.

In DFT, this electron density can be represented by atomic orbital function, as a basis set consisting of radial basis functions and spherical harmonics in three-dimensional space \cite{andzelm1992density, mortensen2005real}. Spherical harmonics are a set of angular basis functions subdivided by a degree ($L$), an integer-valued notation representing an angular frequency of orbitals \cite{morris2005real, mchale2017molecular}. Specifically, $d$ and $f$ orbitals are also called “\textit{polarization functions}”, because they can describe a distortion of an electron cloud, named as polarization \cite{hariharan1973influence, ditchfield1971self, rassolov20016}.

These polarization functions are useful in describing molecular properties including valence electrons, and consequently affect various types of properties of a molecule \cite{helgaker2012recent, karelson1996quantum}. To estimate quantum mechanical properties with considering polarization effects, the basis sets for DFT calculation need to include polarization functions of higher degree \cite{jensen2001polarization, sanchez1997density}. QM9 \cite{ramakrishnan2014quantum}, one of the popular datasets in deep learning benchmark, was also calculated by 6-31G(2df,p) level of basis sets \cite{davidson1986basis} which containing polarization basis.

The previous SO(3)-equivariant networks including Equiformer \cite{liao2022equiformer}, NequIP \cite{batzner20223}, and others \cite{anderson2019cormorant, fuchs2020se, brandstetter2021geometric, unke2021spookynet, batatia2022mace, frank2022so3krates} also used radial basis and spherical harmonics in their networks to represent molecule conformation. However, this approach may be limited because the representation of a set of atoms as points cannot cover the electron density functional. 
Therefore, most existing SO(3)-equivariant networks are potentially limited in prediction of molecular potential energy and related properties, without expression of electron density.

To address these challenges, we propose a novel flexible extension method for SO(3)-equivariant networks motivated by DFT, \textit{Neural Polarization}, by allowing each equivariant block to explicitly consider the polarization effect of electron density while keeping SO(3)-equivariance. The key point of Neural Polarization is introducing an additional “\textit{movable point}”, which is similarly defined as the existing atom, but these points can update their location during the training process. These moving points can be viewed as a type of direction indicator describing the polarized electron density, which is closer to a space-filling view of a molecule. We did not use any additional constraint on movements of these movable points, expecting that atomic orbital polarization caused by electron configuration can be learned for better molecule representation learning.

We applied Neural Polarization on three existing SO(3)-equivariant networks for quantum mechanical property prediction, and trained the extended models from scratch (without pretrained parameters). We verified that Neural Polarization significantly improved the prediction performance over a wide range of targets compared with the original report, especially thermodynamic potential-related targets. We also visualized the trajectories of each movable point in the three trained models, and observed that the shifting patterns of movable points have distinctive characteristics according to the target objectives. The experimental results support our initial assumption that Neural Polarization, which explicitly models the directional surroundings of an atom, induces the latent features to exhibit behavior more similar to the electron density in DFT. We also analyzed the pattern of the position of movable points, comparing with the fundamentals of chemical bonding. Finally, we mathematically verified that equivariant networks equipped with Neural Polarization also have the strictly lower bound of an approximation error with the same maximum degree of spherical harmonics and higher model expressiveness, compared to the original networks.

The contributions of this study are summarized as below.

1. Motivated by DFT, we developed a novel extension method, Neural Polarization, for SO(3)-equivariant networks by introducing movable points expecting that their positions can incorporate the effect of polarization representation for advanced molecular representation.

2. We validated the effectiveness of Neural Polarization based on the performance gain of the experimental results. In addition, the trajectories of movable points from the trained models showed that Neural Polarization can adaptively find the better description of polarization, depending on the target objective.

3. We verified that Neural Polarization lowers the approximation error trained with the same spherical harmonics with the original network, and improves the model expressiveness.

%% file: preliminary.tex
\section{Preliminary of equivariant neural networks}
\label{Preliminary}

In this manuscript, we aim to present the most important part of the preliminary due to space constraints. The remaining parts, DFT and SO(3)-equivariance are introduced in the Appendix \ref{appendix.equivariance}.

By definition, a group equivariant network consists of group equivariant layers such that its output transforms equivariantly under specified group operations applied to its input \cite{scott2012group}. Meanwhile, a group-invariant network has the final layer which is group-invariant, and all other group-equivariant layers. For predicting molecular property $y$ related to its energy for learning molecule embedding representation $E(\mathbf{x})$ from  atom position $\mathbf{x}=\{x_i\}$, the network of $T$ layers should be group-invariant: consisting of $0,...,T-1$ group-equivariant layers $M$ with the final group-invariant readout function $R$, or a pooling layer. That is, $\hat{Y} = R \circ {M_{T-1}} \circ {M_{T-2}} \circ \ldots \circ {M_0} \circ {E}(\mathbf{x})$.

\textbf{Embedding layer.} In general, the embedding layer $E$ locates at the front of the network. The embedding layer learns a feature vector of individual atoms $\mathbf{v}_0=\{v_i\}$ from atom positions $\mathbf{x}=\{x_i\}$ and atom numbers $\mathbf{a}=\{a_i\}$.

\textbf{Equivariant layer.} A function $f$ which satisfies $g \circ f(x) = f(g \circ x)$ for any group element $g \in G$, is called a $G$-equivariant layer. For representing molecule structures in Euclidean space, an orthogonal geometry group SO(3) or SE(3) is generally selected as $G$. Many equivariant networks \cite{thomas2018tensor, schutt2021equivariant, unke2021spookynet, anderson2019cormorant, batatia2022mace} utilized a message-passing function \cite{gilmer2017neural} as a framework for their equivariant layers. However, there is no explicit constraint on the choice of an architecture. For example, \cite{liao2022equiformer, tholke2022torchmd} consists of equivariant self-attention layers for molecules.
We denote that $M_t$ uses given position $\mathbf{x}$ and learnable equivariant vector $\mathbf{v_t}$, however, some networks only uses $\mathbf{v_t}$ for learning $\mathbf{v_{t+1}}$.

\textbf{Pooling (Readout) layer }. A pooling layer $R$ locates the end of a network, also called a readout function, produces the output $\hat{y}$. In molecular property prediction, this layer merges all equivariant feature vectors and produces a scalar-valued prediction. 

\label{eq:layered_architecture}
\begin{align}
    Embedding \quad layer: \mathbf{v}_0 
 &= E(\mathbf{x}, \mathbf{a}) \\
    Equivariant \quad layer: \mathbf{v}_{t+1} &= M_{t}(\mathbf{x}, \mathbf{v}_{t})  \\
	Pooling \quad layer: \hat{y} &= R(\mathbf{v}_{T}) 
\end{align} 

%% file: method_and_experiment.tex
\section{Methods}

\subsection{Motivation of Neural Polarization by DFT}

\textbf{The role of electron density in DFT calculation}

A molecule conformation is represented by a set of atomic number $\mathbf{a}$ and positions $\mathbf{x}$ of constituting $N$ atoms.
Molecular property $\hat{y}$ can be predicted from its conformation $X=\{x_i, a_i\}_{i=1,...,N} = \{\mathbf{x}, \mathbf{a}\}$.
DFT has been most widely used method for addressing this problem, providing more accurate and reliable predictions compared to recent deep learning-based approaches \cite{kalita2021learning, schleder2019dft}.
We hypothesized that the prediction performance of deep neural networks would benefit from fundamental concepts of DFT. In particular, we aim to propose a methodology for improving the prediction performance of existing equivariant networks based on the fundamentals of DFT.

As mentioned in Appendix \ref{appendix.DFT}, DFT calculation is achieved by two sequential steps \cite{engel2011density}. The first step is calculating electron density $\rho$ of from the given $X$, and predict molecular property $\hat{y}$ with the well-defined functional based on the calculated $\rho$, which is a type of function defined on any vector $\vec{r} \in \mathbb{R}^3$. The first step $X \rightarrow \rho$ is achieved by solving computation-intensive Kohn-Sham equation \cite{kohn1965self}, whereas molecular energy (or $\hat{y}$) can be easily calculated from the electron density $\rho$ using pre-defined functionals, in the second step.

\textbf{Construction of a link between electron density and feature space of equivariant networks}

Based on these principles, we developed the assumption that if the latent feature $\chi = \{\mathbf{x}, \mathbf{v}\}$ of any equivariant network is equivalent to $\rho$, or if $\chi$ can express all information contained in $\rho$, the network would suggest more accurate and reliable predictions close to DFT. To be more concrete, our research objective is to develop a novel equivariant latent feature $\chi$ with atom positions $\mathbf{x}$ and $\mathbf{v}$ for SO(3) equivariant networks, of which each layer can approximate the electron density $\rho$ of a given molecule in DFT calculation. 
To be precise, we aim to train $\xi$ and appropriate $\chi$ which satisfies $\xi(\chi) = \rho$.

In DFT calculations, the electron density $\rho$ is expressed as a linear combination of a finite set of basis functions. Analogously, the latent feature in equivariant networks resides in a finite-dimensional vector space. In addition, in DFT, finding the optimal representation for $\rho$ using basis sets is analogous to finding $\xi$ under the constraints of linearity and invertibility. Based on this connection between finding optimal DFT basis sets and constructing an expressive latent space for SO(3)-equivariant networks for molecule property prediction, we aim to introduce the methodology based on selecting the DFT basis set, in order to improve the performance of the neural network.

Among factors considered for basis set, we focused on the term "polarization", which is one of the significant characteristics of a molecule. Polarization refers to a distortion toward specific direction $\tilde{x}$ of the electron cloud depending on electron configuration around an atom nuclei. The shape of polarization is determined by complex interatomic interactions, and has a direct effect on various properties. To incorporate polarization effects, DFT utilizes a high-degree polarization function in general. Analogously, if we learn a basis function for equivariant features $\tilde{v}_i$ of any movable point $\tilde{x}_i$ near the original atom $x_i$, the latent space can effectively learn polarization functions. Therefore, if we can extend an equivariant network to incorporate a pair of ($\tilde{x}_i$, $\tilde{v}_i$) in feature space to get a hint of polarization effect, the network would be more powerful and expressive in representing atom surroundings and show better prediction performance.
In Figure~\ref{fig:overview}, a schematic diagram comparing the concepts is described.

\subsection{Neural Polarization}

 We introduce a high-level description of Neural Polarization, because the internal structure of each module depends on the original baseline networks. Neural Polarization is a type of extension methodology for SO(3)-equivariant networks, with an additional movable point $\tilde{\mathbf{x}}$ of each atom with its corresponding equivariant feature vector $\tilde{\mathbf{v}}$, and $t$ equivariant layers (or blocks) $\tilde{M}_t$ which are extended for incorporating $\tilde{\mathbf{x}}$ and $\tilde{\mathbf{v}}$ as inputs.

The first step is initializing movable points of position $\tilde{\mathbf{x}}_{0}$ and type $\tilde{\mathbf{a}}_0$, and embedding them using $E$ for creating $\tilde{\mathbf{v}}_{0}$. The initial position $\tilde{\mathbf{x}}_{0}$ is same with $\tilde{\mathbf{x}}$ of the original atoms.

\begin{align}
	\tilde{\mathbf{x}}_{0} &= \mathbf{x}\\ 
	\tilde{\mathbf{v}}_{0} &= E(\tilde{\mathbf{x}_0}, \tilde{\mathbf{a}})
\end{align} 

Second, the network updates $\tilde{\mathbf{v}}_{t}$ with  $\tilde{M}_t$. Contrary to an original $M_t(\mathbf{x}_t, \mathbf{v}_t)$,  $\tilde{M}_t$ is defined on extended inputs $([\mathbf{x};\tilde{\mathbf{x}}_t], [\mathbf{v}_t;\tilde{\mathbf{v}}_t])$. Our equivariant $\tilde{M}_t$ updates $(\mathbf{v}_{t+1}, \tilde{\mathbf{v}}_{t+1})$ based on $(\mathbf{x}, \tilde{\mathbf{x}}_{t}, \mathbf{v}_{t}, \tilde{\mathbf{v}}_{t})$ using ${M}_t$. 

\begin{align}
	\tilde{M}_t: [\mathbf{v}_{t+1};\tilde{\mathbf{v}}_{t+1}] = M_t([\mathbf{x};\tilde{\mathbf{x}}_t], [\mathbf{v}_t;\tilde{\mathbf{v}}_t])
\end{align}

$\tilde{\mathbf{x}}_{t+1}$ is produced by the additional projection layer $\pi_{t}: \mathbb{V}^{k\times N} \rightarrow \mathbb{R}^{3\times N} $, given by an equivariant feature $\tilde{\mathbf{v}_t}$ of $\tilde{M}_t$. We constructed a $\pi_{t}$ as a sequential block of an equivariant layer and linear layer, however, there is no constraint on the constitution of $\pi$ block.
Note that $\tilde{M}_t$ does not modify the original atom position $\mathbf{x}$, following the baseline networks.

\begin{align}
	\Delta\tilde{\mathbf{x}}_t = \pi_{t}(\tilde{\mathbf{v}_t})\\ 
    \tilde{\mathbf{x}}_{t+1} = \tilde{\mathbf{x}}_{t} + \Delta\tilde{\mathbf{x}}_{t}
\end{align}

The overview and psuedocode of neural polarization is described in Figure \ref{fig:architecture} and Algorithm~\ref{alg:NP}, respectively, compared with the original framework. In broad terms, the optimizing $M_t$ and $\{x, \tilde{x}, v, \tilde{v}\}$ may correspond to finding the optimal $\xi$ and $\chi$, respectively.

\subsection{Mathematical interpretation of Neural Polarization}

We introduced the process of training movable points and their equivariant features $\{ \tilde{\mathbf{x}}, \tilde{\mathbf{v}} \}$ in networks with Neural Polarization. To investigate the advantage of $\{ \tilde{\mathbf{x}}, \tilde{\mathbf{v}} \}$ in approximating $\rho$, we also conducted theoretical analysis on these terms.
In particular, We will discuss about approximation capability of Neural Polarization for electron density. To discuss the approximation capability for the electron density $\rho$, we introduce the following definition.

\begin{definition}
\label{definition:error}
    Let define the error $\mathcal{E}(\rho, \hat{\rho})$ between electron density $\rho$ and $\hat{\rho}$ as $\mathcal{E}(\rho, \hat{\rho})=\int_{\mathbf{R}^3}{|{(\rho - \hat{\rho})}|^2{dV}}$ and error $P$ and $\rho$ as $\mathcal{E}(P, \rho) = \underset{\rho \in P}{\mathrm{min}} \: \mathcal{E}(\rho, \hat{\rho})$.
\end{definition}

The error $\mathcal{E}(P, \rho)$ defined in Definition~\ref{definition:error} can be regarded as a metric for approximation capability of $P$ about $\rho$. Let $P$ and $\tilde{P}[\tilde{\mathbf{x}}]$ denote the latent feature of electron density in the original network and the network with Neural Polarization, respectively. Then, the following holds, by setting $\tilde{\mathbf{x}}$ where error $\mathcal{E}(P, \rho)$ occurs. Detailed definition and proof are provided in \ref{appendix:proof.NP.approx}.

\begin{proposition}
\label{proposition:expressiveness}
    For any electron density $\rho$, there exists $\tilde{\mathbf{x}}$ which that satisfies $\mathcal{E}(P, \rho) > \mathcal{E}(\tilde{P}[\tilde{\mathbf{x}}], \rho)$ 
\end{proposition}

Proposition~\ref{proposition:expressiveness} shows that Neural Polarization can achieve better approximation than the original network for arbitrary electron densities. Because the electron density itself is a type of function, this proposition supports that Neural Polarization can obtain node features $\mathbf{v}$ that are closer to the real electron density compared to the original network. Therefore, we have demonstrated that Neural Polarization provides better approximation to the electron density within neural networks.

\section{Experiment}

To confirm the effect of Neural Polarization on general SO(3)-equivariant models, we selected three equivariant models (EGNN \cite{satorras2021n}, Equiformer \cite{liao2022equiformer}, TorchMD-NET \cite{tholke2022torchmd}) of various architectures as the baseline networks. We performed experiments on QM9 \cite{ramakrishnan2014quantum} and MD17 \cite{chmiela2019sgdml}, which are most commonly used datasets for molecular property predictions. Lastly, we investigated whether Neural Polarization can be effective on non-molecular tasks involving particle movements, we conducted experiments on the n-body system task proposed by in EGNN. Details of implementations are presented in the Appendix \ref{appendix:implementation}.

%% file: result_table.tex
\begin{figure}
    \centering
    \includegraphics[width=.6\linewidth]{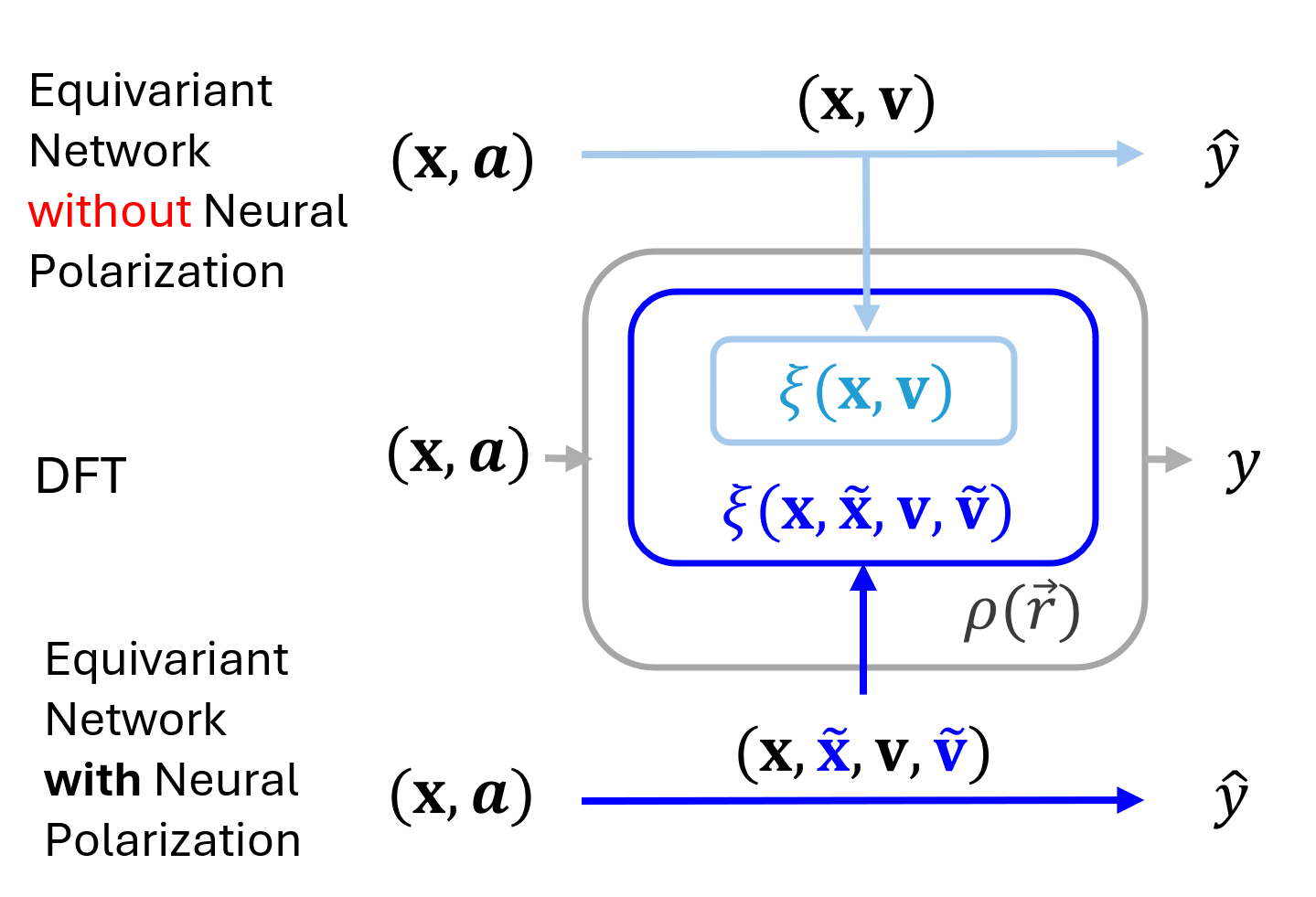}
    \caption{Conceptual overview of our research compared with other methodology. In DFT, there exists a one-to-one correspondence between $\rho(\vec{r})$ and molecular conformation $\{\mathbf{x}, \mathbf{a}\}$ which both fully determines the other properties of the molecule. The baseline $\mathbb{SO}(3)$-equivariant networks can provide a more rich representation of an electron density $\rho(\vec{r})$ with Neural Polarization.}
    \label{fig:overview}
\end{figure}

\begin{figure}
    \centering
    \includegraphics[width=.75\linewidth]{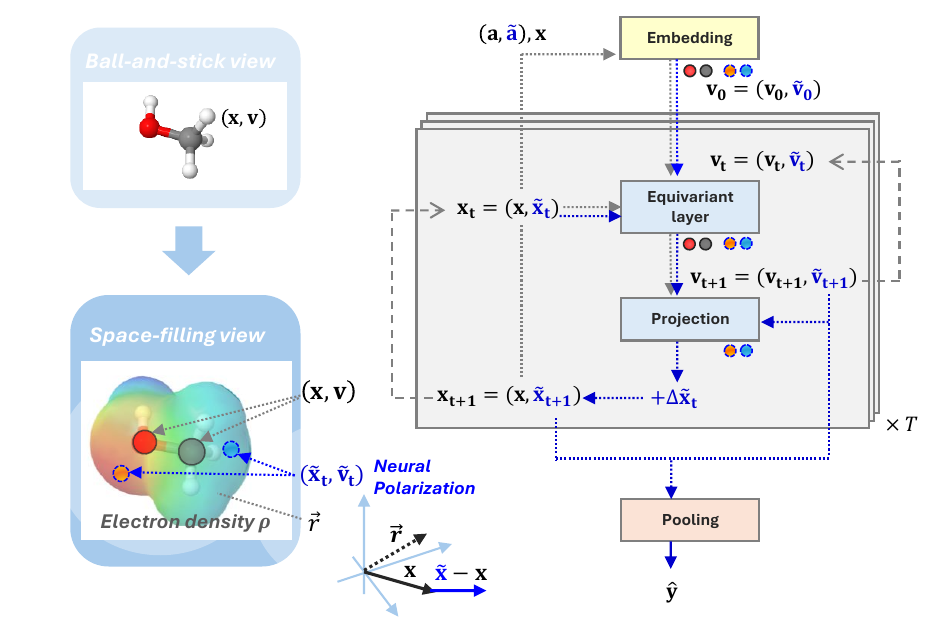}
    \caption{Comparison between ball-and-stick view with space-filling view (left) and overall architecture of equivariant networks accompanied by Neural Polarization (right). Neural polarization allows a molecule representation to incorporate a nearby environment including polarization, which is more similar to a space-filling view of a molecule. We expect two types of moving terms $\tilde{\mathbf{x}}$ and $\tilde{\mathbf{v}}$ can infer and utilize the characteristics of an electron density $\rho$ of each molecule during the training process, without any additional information or constraints.}
    \label{fig:architecture}
\end{figure}

\begin{table}[t]
  \begin{adjustwidth}{-4cm}{-4cm}
  \caption{Mean absolute error on QM9.}
  \centering
  \label{tbl:mae_qm9}
  \begin{tabular}{llccccccc}
    \toprule
    \multicolumn{1}{l}{Target}                   &
    \multicolumn{1}{l}{Unit}                   &
    \multicolumn{2}{c}{EGNN}                   &
    \multicolumn{2}{c}{TorchMD-NET}                   &
    \multicolumn{2}{c}{Equiformer}                   &
    \multicolumn{1}{l}{avg. $\Delta$\%}                   \\
    \cmidrule(lr){3-4}
    \cmidrule(lr){5-6}
    \cmidrule(lr){7-8} &
    \multicolumn{1}{c}{} &
    \multicolumn{1}{c}{w/o NP} &
    \multicolumn{1}{c}{w/ NP} &
    \multicolumn{1}{c}{w/o NP} &
    \multicolumn{1}{c}{w/ NP} &
    \multicolumn{1}{c}{w/o NP} &
    \multicolumn{1}{c}{w/ NP}  \\

    \midrule
    $\mu$ & D & \textbf{0.029} & 0.03 & \textbf{0.011} & 0.014 & 0.011 & \textbf{0.010} & +4.92\% \\
    $\alpha$ & ${a_0}^3$ & \textbf{0.071} & \textbf{0.071} & 0.059 & \textbf{0.0447} & \textbf{0.046} & 0.0527 & \textbf{-3.31\%} \\
    $\epsilon_{\mathrm{HOMO}}$ & meV & \textbf{29} & 29.9 & 20.3 & \textbf{18.4} & \textbf{16.5} & 16.7 & \textbf{-2.04\%} \\
    $\epsilon_{\mathrm{LUMO}}$ & meV & 25 & \textbf{23.4} & \textbf{17.5} & 17.8 & 14.3 & \textbf{14.0} & \textbf{-2.43\%} \\
    $\Delta_\epsilon$ & meV & 48 & \textbf{47.9} & \textbf{36.1} & 41.9 & \textbf{30} &  33.7 & +8.20\% \\
    <$R^2$> & ${a_0}^2$ & 0.106 & \textbf{0.089} & \textbf{0.033} & 0.085 & 0.251 & \textbf{0.162} & \textbf{-4.29\%} \\
    $zpve$ & meV & 1.55 & \textbf{1.50} & 1.84 & \textbf{1.22} & \textbf{1.26} & 2.15  & \textbf{-4.25\%} \\
    $U_0$ & meV & 11 & \textbf{9.52} & 6.15 & \textbf{5.5} & 6.59 & \textbf{5.54} & \textbf{-15.44\%} \\
    $U$ & meV & 12 & \textbf{10.33} & 6.38 & \textbf{5.4} & 6.74 & \textbf{5.49} & \textbf{-19.03\%} \\
    $H$ & meV & 12 & \textbf{9.52} & 6.16 & \textbf{5.6} & 6.63 & \textbf{6.27} & \textbf{-13.93\%} \\
    $G$ & meV & 12 & \textbf{11.5} & 7.62 & \textbf{6.6} & 7.63 & \textbf{7.01} & \textbf{-9.55\%} \\
    $C_v$ & cal/mol K & \textbf{0.031} & 0.032 & 0.026 & \textbf{0.023} & \textbf{0.023} & \textbf{0.023} & \textbf{-3.31\%} \\
    \bottomrule
     &  avg. $\Delta$\%  &  & \textbf{-6.84\%} &  & \textbf{-5.29\%} &  & \textbf{-4.76\%} & \textbf{-5.63\%} \\
    \bottomrule
  \end{tabular}
  \end{adjustwidth}
  \begin{tablenotes}
    \small
    \item NP: Neural Polarization.
  \end{tablenotes}
\end{table} 

\begin{table}
  
  \caption{Mean absolute errors (MAE) of the energy and force prediction of MD17 (Unit: kcal/mol/A).}
  \centering
  \label{tbl:mae_md17}
  \begin{tabular}{llccc}
    \toprule
    \multicolumn{1}{l}{Molecule}                   &
    \multicolumn{1}{l}{Target}                   &
    \multicolumn{2}{c}{TorchMD-NET}                   &
    \multicolumn{1}{l}{avg. $\Delta$\%}                   \\

    \cmidrule(lr){3-4} &
    \multicolumn{1}{c}{} &
    \multicolumn{1}{c}{w/o NP} &
    \multicolumn{1}{c}{w/ NP} \\

    \midrule
    Aspirin & Energy & \textbf{0.123} & 0.126 & 2.38\% \\
                               & Forces & 0.253  & \textbf{0.224} & \textbf{-12.95\%} \\
    Benzene & Energy & 0.058 & \textbf{0.05424} & \textbf{-6.93\%} \\
                              & Forces & 0.196  & \textbf{0.1174} & \textbf{-10.48\%} \\
    Ethanol & Energy & \textbf{0.052} & 0.0524  & 0.76\% \\
                              & Forces & 0.109  & \textbf{0.0878} & \textbf{-24.15\%} \\
    Malonaldehyde & Energy & \textbf{0.077} & 0.0794 & 3.02\% \\
                              & Forces & 0.169  & \textbf{0.146} & \textbf{-15.75\%} \\
    Naphthalene & Energy & 0.085 & \textbf{0.081} & \textbf{-4.94\%} \\
                              & Forces & \textbf{0.061}  & 0.1594  & 61.73\% \\
    Salicylic acid & Energy & 0.093 & \textbf{0.08086} & \textbf{-15.01\%} \\
                              & Forces & 0.129  & \textbf{0.1262} & \textbf{-2.22\%} \\
    Toluene & Energy & 0.074 & \textbf{0.058} & \textbf{-27.59\%} \\
                              & Forces & 0.067  & \textbf{0.057} & \textbf{-17.54\%} \\
    Uracil & Energy & 0.095 & \textbf{0.0857} & \textbf{-10.85\%} \\
                             & Forces & 0.095  & \textbf{0.0857} & \textbf{-10.85\%} \\
    \bottomrule
                            & Energy &        &                  & \textbf{-7.39\%} \\
                             & Forces &        &                  & $^*$\textbf{-4.03\%} \\
    \bottomrule
  \end{tabular}

  \begin{tablenotes}
    \small
    \item *The performance gain is -13.42\% except for the case of Naphthalene.
  \end{tablenotes}
\end{table}


\newsavebox{\algleft}
\newsavebox{\algright}

\savebox{\algleft}{%
\begin{minipage}{.49\textwidth}
  \begin{algorithm}[H]
    \caption{SO(3)-equivariant network without Neural Polarization}
    \begin{algorithmic}
      \scriptsize
      \STATE Given $ \mathbf{x} \in \mathbb{R}^3 $, $ \mathbf{a} \in \mathbb{R}$, $ \mathbf{v} \in \mathbb{V}$ and a layer index $t=0,1,...,(T-1)$.
      \STATE $\mathbf{v_0}$ $\longleftarrow$ Embedding($\mathbf{x}$, $\mathbf{a}$)
      
      \FOR{$ t=0,1,...,(T-1) $}
        \STATE $\mathbf{v}_{t+1}$ $\longleftarrow$ EquivariantLayer($\mathbf{v}_{t}$)      
      \ENDFOR
      
      \STATE {$\hat{y}$ $\longleftarrow$ Pooling($\mathbf{v}_T$)}
      \STATE {return $\hat{y}$}
      \\
      \\
      \\
    \end{algorithmic}
  \end{algorithm}%
\end{minipage}}%

\savebox{\algright}{%
\begin{minipage}{.49\textwidth}
  \begin{algorithm}[H]
    \caption{SO(3)-equivariant network with Neural Polarization (proposed)}
    \label{alg:NP}
    \begin{algorithmic}
      \scriptsize
      
      \STATE Given $ \mathbf{x}, \color{blue}{\mathbf{\tilde{x}}}$ $ \in \mathbb{R}^3 $, $ \mathbf{a}, \color{blue}{\mathbf{\tilde{a}}} \in \mathbb{R}$, $ \mathbf{v}, \color{blue}{\mathbf{\tilde{v}}} \in \mathbb{V}$ and a layer index $t=0,1,...,(T-1)$.
      \STATE $\mathbf{v_0}, \color{blue}{\mathbf{\tilde{v_0}}}$ $\longleftarrow$ Embedding($\mathbf{x}$, $\mathbf{a}, \color{blue}{\mathbf{\tilde{a}}}$)
      
      \FOR{$ t=0,1,...,(T-1) $}
        \STATE $\mathbf{v}_{t+1}, \color{blue}{\mathbf{\tilde{v}}_{t+1}}$ $\longleftarrow$ EquivariantLayer([$\mathbf{x}_t, \color{blue}{\mathbf{\tilde{x}}_t}$], [$\mathbf{v}_t, \color{blue}{\mathbf{\tilde{v}}_t}$])      
      
        \STATE $\color{blue}{\Delta \mathbf{\tilde{x}}_t}$ $\longleftarrow$ $\color{blue}{\text{Proj}_t}$ $(\color{blue}{\mathbf{\tilde{v}}_{t+1}}$$)$
        \STATE $\color{blue}{\mathbf{\tilde{x}}_{t+1}}$ $\longleftarrow$ $\color{blue}{\mathbf{\tilde{x}}_{t}}$ + $\color{blue}{\Delta \mathbf{\tilde{x}}_t}$
      \ENDFOR
      \STATE {$\hat{y}$ $\longleftarrow$ Pooling($\color{blue}{\mathbf{\tilde{v}}_T}$)}
      \STATE {return $\hat{y}$}

    \end{algorithmic}
  \end{algorithm}
\end{minipage}}%

\noindent\usebox{\algleft}\hfill\usebox{\algright}%

\begin{figure}
\centering
    \includegraphics[width=.95\linewidth]{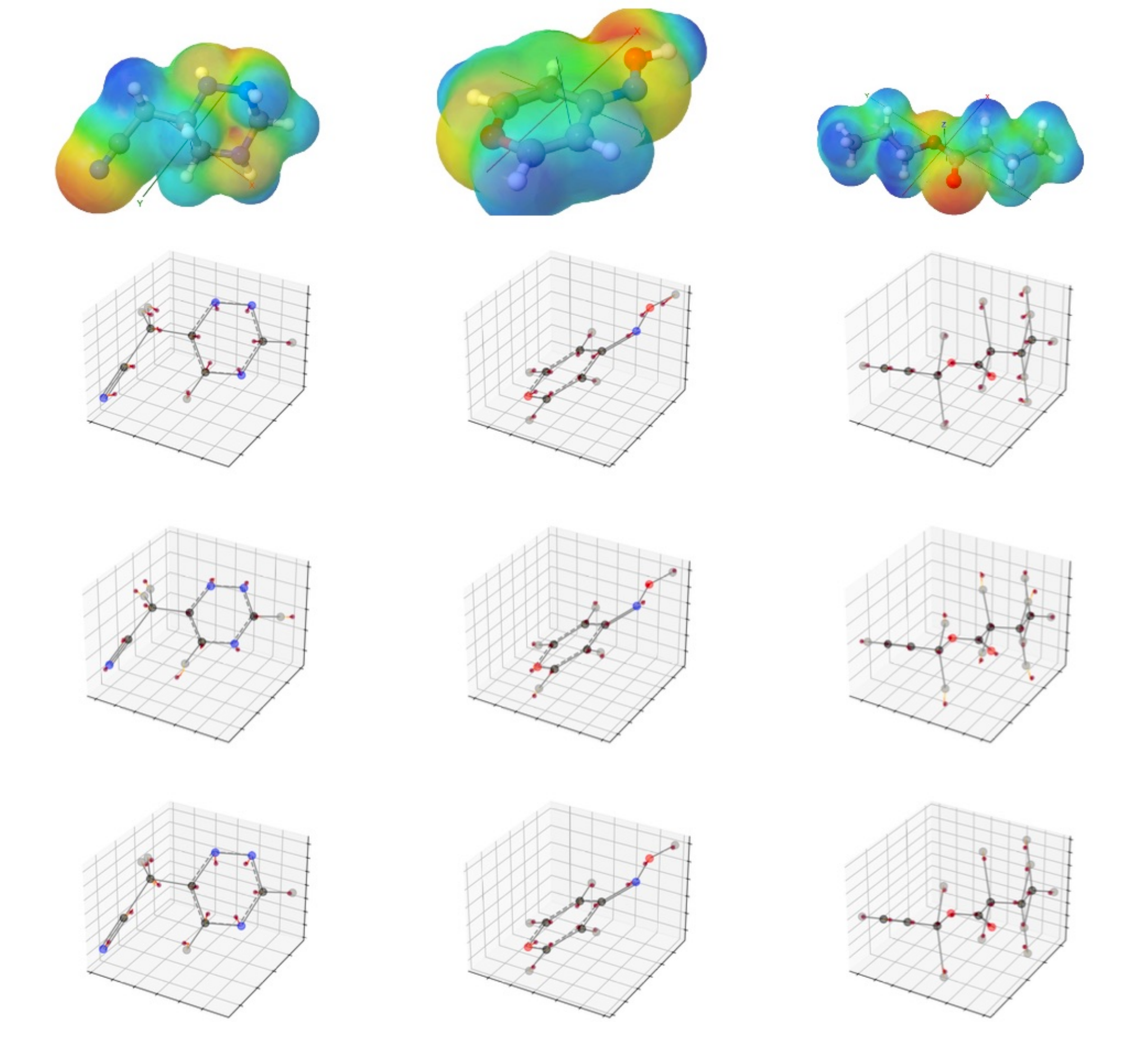}
    \caption{The trajectories of $\tilde{x}$ of three different molecules trained on three types of target objectives, $\epsilon_{HOMO}$ (top), $zpve$ (middle), and $U_0$ (bottom). The first row is the distribution of electronegativity of each target molecule, calculated by GAMESS \cite{GAMESS} the level of 6-31G. The second, third, and the last rows show the trajectories 
    of $\tilde{x}$ (small dark red dots near each atom) from three different models trained on $\epsilon_{HOMO}$, $zpve$, and $U_0$, respectively. The baseline network was torchMD-NET. More figures are provided in \ref{appendix.traj}}
    \label{fig:app.hlumo}
\end{figure}

\begin{table}
  \caption{Ablation results trained with Equiformer on three types of QM9 targets $\mu$, $\epsilon_{HOMO}$ and $U_0$. 
  $\{\mathbf{x}, \mathbf{x}\}$ in the second row is the ablation study for comparing the effect of $\tilde{\mathbf{x}}$, keeping the same computational cost and weight parameters. Scale is the same with Table~\ref{tbl:mae_qm9}.}
  \centering
  \label{tbl:ablation}
  \begin{tabular}{lccc}
    \toprule
    Method &  $\mu$ & $\epsilon_{HOMO}$ & $U_0$ 
 \\
    \midrule
    {Equiformer} & 0.0118 & \textbf{16.5} & 6.59 \\
    {Equiformer + $\{\mathbf{x}, \mathbf{x}\}$} & 0.0159 & 17.7 & 8.80 \\
    {Equiformer + NP} & \textbf{0.0109} &  16.7 & \textbf{5.54} \\
    \bottomrule
  \end{tabular}
\end{table}

\begin{table}
  \caption{Mean Squared Error (MSE) for the future position estimation in n-body task (the prediction of particles' movement), proposed in \cite{satorras2021n}. The results of baseline EGNN are retrieved from the original paper.}
  \centering
  \label{tbl:nbody}
  \begin{tabular}{lc}
    \toprule
    Method & MSE \\
    \midrule
    {EGNN} & 0.0071 \\
    {EGNN + NP} & \textbf{0.0051} \\
    \bottomrule
  \end{tabular}
\end{table}

%% file: result.tex
\section{Result}

\subsection{The performance gains on QM9 and MD17 dataset}

The experimental results on the QM9 dataset are presented in Table~\ref{tbl:mae_qm9}, categorized by whether Neural Polarization was applied (marked as `with NP') or not for each baseline network, in all 12 target cases. Most of the baseline results (the left side) of each previous model were reproduced by training the source code provided in the official page, from the scratch. A few cases were retrieved from the reports on the original paper, in the case of a computation or compatibility issue with our environment for the source code. 

In the case of EGNN, we observed that the error was reduced on 9 labels (including no change of an alpha case) with using a Neural Polarization, with an average of -6.84\% error change rate of all 12 targets.  
In the next case, TorchMD-NET, the error was reduced on 7 targets with a Neural Polarization, with an average of -5.29\% error change rate of all 12 targets. The last case Equiformer, the error was reduced on 8 targets with a Neural Polarization (including no change of a $C_v$ case), with an average of -4.76\% error change rate of all targets.

Interestingly, we observed that for the cases of thermodynamic properties including $U_0, U, H, G$, the error was significantly reduced regardless of the baseline network type. In these four cases, the average performance gain (the average of three error change rates of each baseline case) the ranges from -10\% to -20\%, whereas other six target cases resulting the error reduction (except for $\mu$ and $\Delta_\epsilon$), showed the ranges from -2\% to -5\% of performance gains, respectively.
In the analysis of the average error change rate of each baseline model, there was no significant difference between three models. EGNN showed the best performance gain of -6.84\%, followed by TorchMD-NET (-5.29\%), and Equiformer (-4.76\%).

Next is the analysis on the MD17 dataset consisting of eight molecules, using TorchMD-NET as a baseline network. Each molecule has two types of targets, energy and forces, respectively. These results are presented in Table~\ref{tbl:mae_md17}. In case of energy prediction, the error was reduced in five molecules when trained on the network with Neural Polarization, and the average error change rate of all energies of eight molecules is -7.39\%. In the case of forces prediction, the error was reduced seven of eight molecules on the network with Neural Polarization, except for Naphthalene. We observed that in the case of force prediction of Naphthalene molecule, the error was significantly increased by 61.73\%, although the error was decreased in the case of energy prediction. The reason of this contrasting results of the naphthalene case is not clear. In summary, the average error change rates were -7.39\% in energy predictions and -4.03\% in forces predictions of eight molecules, respectively. Without considering naphthalene, the average error change rate of force predictions was decrease to -13.42\%.

\subsection{Investigation of the polarization trajectory}

To validate our assumption that updating $\tilde{\mathbf{x}}$ and its equivariant feature $\tilde{\mathbf{v}}$ can facilitate exploiting molecule's electron density for molecular property prediction, we analyzed the final position of $\tilde{\mathbf{x}}$ extracted from the trained models. We selected targets of various types, and tracked the trajectory of every $\tilde{\mathbf{x}}$ during the training process.

We observed that the most determining factor of the movement of $\tilde{x}$ is the target objective, rather than atom type or bond types. $\tilde{x}$ from the model trained on $\epsilon_{homo}$  whereas in the trained model with Neural Polarization on $zpve$, $\tilde{x}$ tends to move toward the outside of the molecule center of mass.
One possible explanation for these characteristic patterns is that Neural Polarization was adaptively trained for optimizing $\tilde{x}$ depending on training target types, rather than just increasing number of parameters for the original atoms.

Another notable point is that each $\tilde{x}$ did not deviate more than a half of the bond length from its belonging atom $x$. Although further profound analyses would be needed to explain the movement of $\tilde{x}$, this trend may be one of the evidences that $\tilde{x}$ perform a role in supporting original ${x}$, while understanding the characteristics of molecular properties. 

\subsection{Ablation study}

We assumed that the performance improvement in the model is not simply caused by an increase in the number of variables and parameters, we trained Neural Polarization on a pair of non-movable points $\{\mathbf{x},\mathbf{x}\}$, which is a replicating the inputs. We conducted an ablation study on $\epsilon_{HOMO}$, $\mu$, and $U_0$ in QM9. As shown in Table~\ref{tbl:ablation}, using $\{\mathbf{x},\mathbf{x}\}$ rather than $\{\mathbf{x},\tilde{\mathbf{x}}\}$ increase the prediction error, and the performances were improved only applied with Neural Polarization. Based on this, we found that movable points perform a significant role in improving performance of property prediction tasks.

\subsection{Neural Polarization in other domain}

We found that \ref{proposition:expressiveness} holds not only for the electron density of molecules but also for general 3-D density functions. 
To examine this assumption, we conducted experiments on the n-body task proposed in \cite{satorras2021n}. As demonstrated in Table~\ref{tbl:nbody}, Neural Polarization also improved performances n-body task, which is not limited to molecule tasks. 
This observation led to possible generalizability of Neural Polarization beyond molecular tasks.

%% file: discussion.tex
\section{Discussion}


For analyzing the effect of Neural Polarization on molecular property prediction tasks, 
we analyzed the final positions of $\tilde{x}$ along with the directions of the covalent bonds in molecule. According to chemistry, a single bond is formed by the sharing of an electron pair between two atoms, while a double bond arises from the sharing of two pairs of electrons, leading to electron densities aligned parallel to the bond axis. In case of aromatic rings, the delocalized electron densities form planar regions above and below the ring plane. In accord with this fundamentals, as shown in Figure~\ref{fig:app.hlumo} (right), the trajectories from atom with single bonds exist near bonds, while the trajectories from aromatic rings were created on the same plane with the ring, as shown in Figure~\ref{fig:app.hlumo} (left). In addition, we observed that most of the final location of $\tilde{\mathbf{x}}_{T}$ (small red dot) is distant from the original atom location $\mathbf{x}_{0}$, but no more than half the bond length away, which is related with the inherent property of a covalent bond. 
From those observations, we assume that Neural Polarization recognize the various molecular property including the characteristics of bonding types, and understand the characteristics of a polarization effect of each molecule.


%% file: conclusion_and_limitation.tex
\section{Conclusion}

We proposed Neural Polarization, which enables the intermediate state of an equivariant network to better represent the electron density corresponding to the intermediate state in DFT calculations. Accompanied by flexible applicability, Neural Polarization demonstrated performance improvements across diverse tasks and models, showing that it can be trained toward the polarization characteristics of electron density in accord with our assumption. Based on these results, we expect Neural Polarization to enforce improvements in general molecular problems. For future work, we will propose various methodologies inspired by more aspects beyond polarization.

\section*{Limitations}
While Neural Polarization does not change the computational complexity of the original model, it introduces an additional computational cost. 
Meanwhile, for deeper insights from trajectories, it would be beneficial to validate the approach on molecular datasets that include electron density information.

\section*{Broader impacts}
Our study can lead to an advanced research subjects for bridging gap between quantum chemistry and deep learning.
In addition, the development of Neural Polarization involves concepts from DFT, equivariant neural networks, and molecular modeling. This interdisciplinary approach could foster collaborations between researchers from different fields, such as physics, chemistry, machine learning.

%% file: related_work.tex
\subsection{Table of notations}

\input{variable_table}

\subsection{Related works}

\textbf{Molecule property prediction based on molecule structure in Euclidean space}
Based on the relationship between molecule conformation and its properties, many existing networks utilized molecule conformation as a source for property prediction tasks. SchNet \cite{schutt2017schnet} introduced a radial basis function for embedding continuous-valued atom-atom distances, and DimeNet \cite{klicpera_dimenetpp_2020} utilized Bessel basis functions for embedding continuous-valued angles between three atoms.

To more accurately represent molecule structures located in Euclidean space, recent studies introduced geometry-group theory in their network. Various types of representation theory have been utilized in modern chemistry for structural analysis of molecules or crystals. Cormorant \cite{anderson2019cormorant}, NequIP \cite{batzner20223}, GemNet \cite{gasteiger2021gemnet}, PaiNN \cite{schutt2021equivariant}, and SpookyNet \cite{unke2021spookynet} are the well-known examples of their geometry-group equivariant blocks for molecule conformation. These studies contributed to more accurate and reliable molecule structure learning, however, the effects caused by electron configurations may still be limited in these type of representation.

\textbf{Two branches of implementing equivariant networks}
Broadly speaking, there are two branches for implementing geometry-group equivariance in a neural network. One branch relied on the representation technique for SO(3) group in Physics. In particular, they introduced Clebsch-Gordan coefficients \cite{de2018octet} or Wigner-D matrices \cite{wigner1931gruppentheorie}, for SO(3)-group equivariant tensor product for features defined on spherical basis. Cormorant \cite{anderson2019cormorant}, NequIP \cite{batzner20223}, SE(3)-Transformer \cite{fuchs2020se} are the examples belong to this category. Further explanation is described in \ref{appendix.equivariance}.

On the other hand, EGNN \cite{satorras2021n} and several following works \cite{deng2021vector, hoogeboom2022equivariant} did not use a computationally expensive tensor production. Instead, they separated scalar-valued features for scales and vector valued-features for directions, and trained them as individual features for molecule structure. This approach is relatively efficient in terms of computational complexity for larger molecules, in general.

\textbf{Utilization of DFT for property prediction in machine learning}
As machine learning-based methods have progressed for solving more sophisticated problems in chemistry and material science, many recent studies \cite{ryczko2019deep, schutt2019unifying, kalita2021learning, pederson2022machine, huang2023central} focused on DFT for a wide range of tasks. \cite{xie2018crystal, allam2018application} studied the DFT-related property prediction for the given materials. \cite{lee2022machine, chen2022improving, huang2022provably} are the examples of considering DFT for other molecules-related tasks. In one of the review paper \cite{huang2023towards}, the authors argued that understanding DFT will be the necessary background to explore the chemical property for machine learning-based methods.

The prediction performance of current networks
Despite the rapid increase in prediction performance of deep neural networks over a short period, there remains a considerable gap compared to DFT methods. There have been several reports \cite{duan2021putting, levine2014quantum, chmiela2019sgdml, yu2024qh9} about these limitations of the current neural networks in Chemistry research, and they pointed out that one of the possible limitations is that the deep learning approaches could not utilize enough chemical information in appropriate ways, including DFT.

%% file: variable_table.tex
\begin{table}[h]
  
  \caption{Table of notations in this manuscript}
  \centering
  \label{tbl:variables}
  \begin{tabular}{cc}
    \toprule
    Variable & Definition
 \\
    \midrule
    $i$ & an index of an atom of a molecule \\
    $t$ & a layer index of a baseline network \\
    $N$ & the number of atoms in a molecule \\
    $x_i$ & a three-dimensional coordinate of $i$-th atom \\
    $a_i$ & an atom number (or atom type) of $i$-th atom \\
    $v_{t, i}$ & a node feature of $t$ - th Layer output \\
    $\mathbf{x}$ & a set of $\{x_0, x_1, ... , x_{N-1}\}$ \\
    $\hat{y}$ & predicted target (molecular property) \\
    $X$ & molecule conformation \\
    $\rho$ & electron density \\
    $\chi$ & latent feature of a baseline network \\
    $\xi$ & mapping from latent feature $\chi$ to electron density $\rho$ \\
    \bottomrule
  \end{tabular}
\end{table}

%% file: dft.tex
\subsection{Brief introduction about DFT}
\label{appendix.DFT}

Density functional theory (DFT) \cite{levine2014quantum} is one of the most widely used methodologies for studying molecular properties. The overall process of DFT calculation consists of inferring the electron density and calculating the properties of the molecule based on the computed electron density. Molecules are composed of multiple atoms with surrounding electrons. According to quantum mechanics, the location of an electron cannot be specified as a point, but rather as a probability distribution over the space, called electron density. 
The first Hohenberg-Kohn theorem \cite{hohenberg1964inhomogeneous} states that the ground state electron density of a molecule uniquely determines the external potential, and consequently all ground state molecular properties. This theorem implies that knowing just the electron density is enough to calculate any ground state property including energy.

%% file: equivariance.tex
\subsection{Equivariance and representations in SO(3) and SE(3) symmetry}
\label{appendix.equivariance}

We briefly review several concepts on equivariance as an essential background for our motivation and strategy. 
A group \cite{scott2012group} $(G, \circ)$ is a type of an algebraic structure consisting of a non-empty set $G=\{ g \}$ and a binary operation $\circ: G \times G \rightarrow G$ with satisfying three requirements: an associativity, an identity, and an inverse element. A (left) group action $a$ of $G$ on a set $X$ is a function $a: G \times X \rightarrow X$ with satisfying identity and compatibility for all $g, h \in G$ and all $ x \in X$. 

Group representation \cite{serre1977linear, inui2012group} is $\varphi$ a group homomorphism from a group $G$ to a general linear group $GL(V)$, which enables group actions to be represented as a matrix multiplication in (finite) vector space $V$. If $\varphi : G \rightarrow GL(V)$ has only trivial subrepresentations, it is called an \textit{irreducible representation} or \textit{irrep}. One important property of irreducible representations is the Great Orthogonality Theorem \cite{bhlitem93362} stated by \textit{Schur's orthogonality relations} \cite{hall2013lie}, $\sum^{\{|G|\}}_{R\in G}\varphi^{(L)}(g)_{l_m,l_n}=0$ for $l_n,l_m=1,...,L$ are the dimension of $\varphi, \forall \varphi \neq I_L$. This theorem is proved by \textit{Schur's lemma}, 1) If $V$ and $W$ are not isomorphic, then there are no nontrivial $G$-linear maps between them, and 2) if $V=W$ and $\varphi_V = \varphi_W$, then the only nontrivial $G$-linear maps are the scalar multiplication of the identity.

Any function $f$ satisfies $f (g \circ x) = f(x)$ is called an \textit{invariant} function of group $G$ on $X$, while it is called \textit{equivariant} if it satisfies $f (g \circ x) = g\circ f(x)$.

In this study, we focus on the special orthogonal group $\mathbb{SO}(3)$, the group of all rotations under function composition in three-dimensional Euclidean space. 
The irreducible representations of $\mathbb{SO}(3)$ are called Wigner-D matrices $D^L(g)$ of dimension $2L+1$, and there are $(2L+1) \times (2L+1)$ type of irreps matrices $D^L_{m,m'}$, with $-L \leq m, m' \leq L$, respectively (footnote: we assume integer L, m=m' for real-valued spherical harmonics).

Spherical harmonics \cite{muller2006spherical} is a set of orthonormal basis functions for irreducible representations of $\mathbb{SO}(3)$, and denoted by $Y_m^L(\theta, \phi)$ with an integer degree $l$. On the unit sphere $S^2$, any square-integrable function $s: S^2 \rightarrow \mathbb{C}$ can be expanded as a linear combination $s(\theta, \phi)=\sum_{l=0}^{\infty} \sum_{m=-L}^L f_m^{L\ast} Y_m^L (\theta, \phi)$. Accordingly, any group action $g \in \mathbb{SO}(3)$ can be expressed as a direct sum of Wigner-D matrices $D^L_{m,m'}$ with a change of basis $P: P^{-1}D(g)P=D^\prime (g)$ as follows:
\begin{equation}
    D(g) = P^{-1}\Bigg( \bigoplus_{i}D_{l_i}(g)\Bigg)P = P^{-1}   
    \begin{bmatrix}
    D^{l_0}(g) & & \\
    & D^{l_1}(g) & \\
    & & \ldots
  \end{bmatrix}
  P
\end{equation}

We can write a rotation $R$ in three-dimensional Euclidean space followed as:

\begin{equation}
    Y_m^L(\theta + \Delta\theta, \phi + \Delta\phi) = 
     \sum_{m^\prime=-L}^L \Big[ D^L_{m,m^\prime}(\alpha, \beta, \gamma) \Big]Y^L_{m^\prime}(\theta, \phi)
\end{equation}


For a tensor production of two spherical tensors $f^{L_1}_{m_1}$ and $f^{L_2}_{m_2}$ in $\mathbb{SO}(3)$, the Clebsch-Gordan coefficients \cite{hall2013lie} $C_{(L_1,m_1),(L_2,m_2)}^{(L_3,m_3)}$ are used to assign the numerical value according to each decomposed type of two spherical tensors followed as:

\begin{equation}
\label{eq:equivariant_space}
    f^{L_3}_{m_3} = \sum_{m_1=-L_1}^{L_1}\sum_{m_2=-L_2}^{L_2}C_{(L_1,m_1),(L_2,m_2)}^{(L_3,m_3)}f^{L_1}_{m_1}f^{L_2}_{m_2}
\end{equation}

In quantum chemistry, we can describe an electron charge distribution (cloud) of an atom generated by the interactions between an atom and its electrons \cite{levine2014quantum}. It is called as an \textit{atomic orbital}, and described as spherical coordinates with a radial term $R(r)$ and spherical harmonics $Y_m^l(\theta, \phi)$ of polar angle $\theta$ and azimutal angle $\phi$, with different degree $l$ and order $m$. Integer-valued degree of real spherical harmonics (footnote: notation) $l=0,1,2,...$ correspond to $s, p, d,...$ orbital of an atom, and more higher $l$-basis orbital can capture higher angular frequency of a function $f_m^{L}$ defined on the surface of a sphere $S^2$. The total angular momentum coupling can be described by an element of rotation group $g \in \mathbb{SO}(3)$, and Wigner-D matrix \cite{shiraishispin} $D^L_{m,m^\prime}$ with Clebsch-Gordan coefficients $\{C_{(L_1,m_1),(L_2,m_2)}^{(L_3,m_3)}\}$ are used to product two angular momenta represented as spherical tensors of dimension $2L_1+1$ and $2L_2+1$, respectively.

%% file: other.tex
\subsection{Equivalence between selection of basis set and node feature}

Let assume some basis set of DFT $B$ with spherical harmonics.
Since node feature  $\mathbf{v}_{t} \in \mathbb{V}_{eq}$ computed by equivariant layer is  equivariant under $\mathbb{SO}(3)$ transformation, $\mathbf{v}_{t}$ may available to represented as combination of irreducible representation of  $\mathbb{SO}(3)$. Using notation defined in \ref{eq:equivariant_space} , let us represent $\{L, m\}$-degree $\mathbb{SO}(3)$ irreducible feature of  ${v}_{t,i}$ as $f^{L}_{mi} = \sum_{j} f^{L}_{mij}$. With this definition  we can represent ${v}_{t,i} = \sum_{L,m,i,j} {f^{L}_{mij}}$ .

Now, consider linear map $\xi'_{1/2}$ which maps $f^{L}_{mij}$ into $R_{ijL}(r - x_i)Y^{L}_{m}(\theta_i, \phi_i)$.
Then map $\xi'(\mathbf{v}) =  {\xi'_{1/2}(\mathbf{v})}^2 \in (\mathbb{V}_{eq}^{N} \rightarrow (\mathbb{R}^3 \rightarrow \mathbb{R})) $ maps layer output $\mathbf{v}_t$ into electron density, which can be represented as quadratic form under $B$.
Meanwhile, any electron density yielded from DFT with $B$ also forms quadratic form under $B$. Therefore, any invertible linear map $\xi$ defined for $\mathbf{v}_t$ can be converted into linear map, which its image is subset of electron density that can be generated by DFT by $\xi'_{1/2}\cdot \xi^{-1}$.

\label{appendix:proof.NP.approx}
\subsection{Proof of Proposition~\ref{proposition:expressiveness}}

Let $B_P$ basis set of $P$ and $\tilde{B}_P[\tilde{\mathbf{x}}]$ as basis set of $\tilde{P}[\tilde{\mathbf{x}}]$.
Since basis $B_P$ is equivalent with DFT by equivariant constraint, any basis function in  $B_P$ can be represented by  combination of spherical harmonics and radial function centered at a position $x_i$.

\begin{align}
    B_P = \{ R_{ijL}(r - x_i)Y^{L}_{m}(\theta_i, \phi_i) | 1 \leq i \leq N, -L \leq m \leq L, j \leq J_{Lm} \}
\end{align}

Similarly, a basis set with Neural Polarization $\tilde{B}_P[\tilde{\mathbf{x}}]$ can be represented as below.

\begin{align}
\tilde{B}_{P}[\tilde{\mathbf{x}}] = P \cap \{ R_{ijLm}(r - \tilde{x}_i)Y^{L}_{m}(\tilde{\theta_i}, \tilde{\phi_i}) | i \leq N, -L \leq m \leq L, j \leq J_{Lm} \}
\end{align}

Since $B_P$ is a finite basis, there exists density $\rho$ such that $\rho \notin P=span(B_P)$. 
With this $\rho$, let  $\rho_{min}$ an electron density which yield minimal error $ \mathcal{E}(P, \rho) =  \mathcal{E}(\rho_{min}, \rho)$. When $L=0$, spherical harmonics is constant and $R_{ij0}(r-\tilde{x}_i) \in \tilde{B}_{P}[\tilde{\mathbf{x}}]$.  For convenience let define $\phi[\tilde{x}_i] = R_{ij0}(r-\tilde{x}_i)$.

Since $\mathcal{E}(\rho_{min}, \rho) > 0$ , $\rho - ~\rho \neq 0$ and there exists some $\tilde{x}_i \in \mathbb{R}^3$ such that$\int_{\mathbf{R}^3}(\rho-\rho_{min})\phi[\tilde{x}_i]{dV} = \Delta \neq 0$ because of radial symmetry in $\phi[\tilde{x}_i]$. With this $\tilde{x}_i$ ,
 \begin{align}
    \mathcal{E}(\rho_{min}+\frac{\Delta}{2} \phi[\tilde{x}_i], \rho) &= \int_{\mathbf{R}^3}(\rho-\rho_{min} - \frac{\Delta}{2}\phi[\tilde{x}_i])^2\phi[\tilde{x}_i]{dV} \\
    &= \int_{\mathbf{R}^3}(\rho-\rho_{min})^2 + (\frac{\Delta}{2}\phi[\tilde{x}_i])^2 - \Delta\phi[\tilde{x}_i](\rho-\rho_{min}){dV} \\
    &= \mathcal{E}(\rho_{min}, \rho) + (\frac{\Delta}{2})^2 - \Delta^2 \\
    &= \mathcal{E}(\rho_{min}, \rho) - \frac{3}{4}\Delta^2 \\
    &< \mathcal{E}(\rho_{min}, \rho)\\
\end{align}

Because $\rho_{min} +\frac{\Delta}{2} \phi_{ij0} \in \tilde{P}[\tilde{\mathbf{x}}]$, with $\tilde{x}_i \in \tilde{\mathbf{x}}$
\begin{align}
    \mathcal{E}(\tilde{P}[\tilde{\mathbf{x}}], \rho) \leq \mathcal{E}(\rho_{min}+\frac{\Delta}{2} \phi_{ij0}, \rho) < \mathcal{E}(\rho_{min}, \rho) = \mathcal{E}(P, \rho)
\end{align}

\subsection{Datasets}

The QM9 dataset \cite{ramakrishnan2014quantum} is a dataset containing various molecular properties for 134k small molecules. The molecules included in the dataset are composed of C, N, O, H, and F and contain up to 29 atoms. This dataset is frequently used for measuring performance of deep learning models for molecules. The dataset consists of molecular conformations, the atomic coordinates of the given molecule, and 12 different molecular properties calculated from these conformations, such as internal Energy and dipole moment. 

The MD17 dataset \cite{chmiela2017machine} is a dataset about molecular dynamics (MD) trajectories of 10 small molecules. The dataset consists of molecular conformations from the trajectories with the corresponding molecular energy and atomic forces. The task for this dataset is to predict the forces and energy for a given conformation. 
In this paper, we used the MD17 dataset configuration addressed in TorchMD-NET, which is the base model for comparison.

The n-body system task we experimented was suggested from EGNN, which extended task proposed in \cite{kipf2018neural}. The dataset consists of trajectories generated by five charged particles, whose trajectories are simulated based on charged interactions. The task for this dataset is to predict the positions after 1000 steps from given initial positions.

\subsection{Detailed Implementation of Neural Polarization}
\label{appendix:implementation}

We implemented Neural Polarization based on the official GitHub repositories provided by each model. Neural Polarization was implemented using the same way with three types of baseline networks, except for the projection layer.
For EGNN we used Multilayer Perceptron (MLP) with a single hidden layer of 128 dimension as a node feature. In TorchMD-NET, the projection layer was implemented as a single linear layer without bias for the $l=1$ feature of each layer output $\mathbf{\tilde{v}_{t, i}}$. For Equiformer, the projection layer was implemented as a shallow equivariant network based on a tensor product with one $l=1$ feature.

To minimize undesired effect on performances arose by hyperparameter optimization, we fixed all hyperparameters except the learning rate and the batch size provided in the official repositories. We modified batch size as half only in cases where out-of-memory (OOM) errors occurred. For other implementations such as data splitting, optimizer, learning rate scheduling strategy, and objective loss function, we followed configurations implemented in the official repositories. Experiments were conducted on NVIDIA V100, A40, or A100 GPUs. Each experiment with the QM9 dataset has required maximum 240 GPU hours each, while other experiments required less than 24 GPU hours. Training times differ from model types, because we followed the early-stopping condition implemented in each original repository.

\newpage
\subsection{Trajectories generated by Neural Polarization about molecules in QM9 dataset}
\label{appendix.traj}

\begin{figure}[h]
    \includegraphics[width=0.92\textwidth]
    {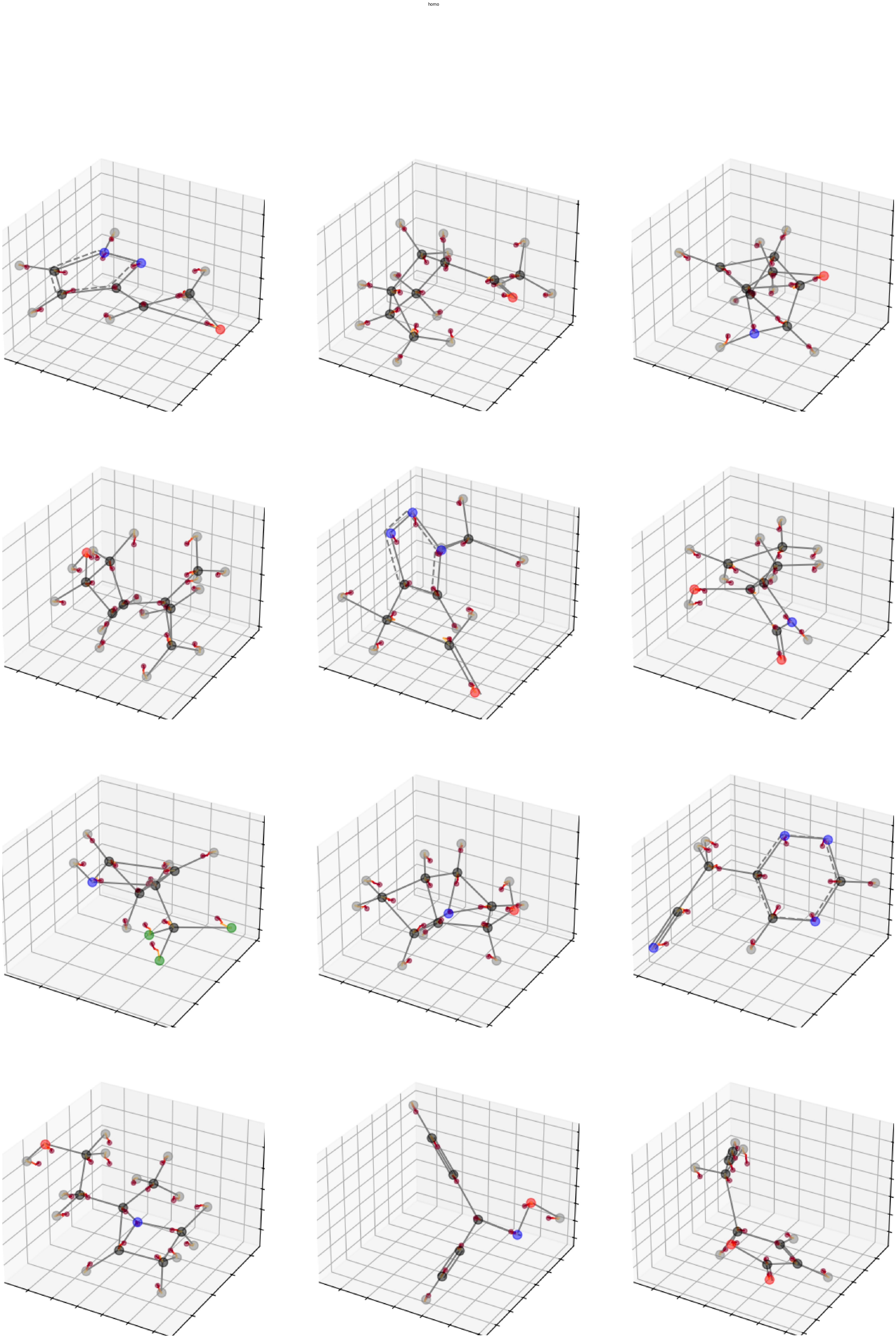}
    \caption{More trajectories of $\tilde{x}$ trained on $\epsilon_{HOMO}$, based on  torchMD-NET.
}
    \label{fig:homo-more-example}
\end{figure}

\newpage

\begin{figure}[hp]
    \centering
    \includegraphics[width=0.92\textwidth]{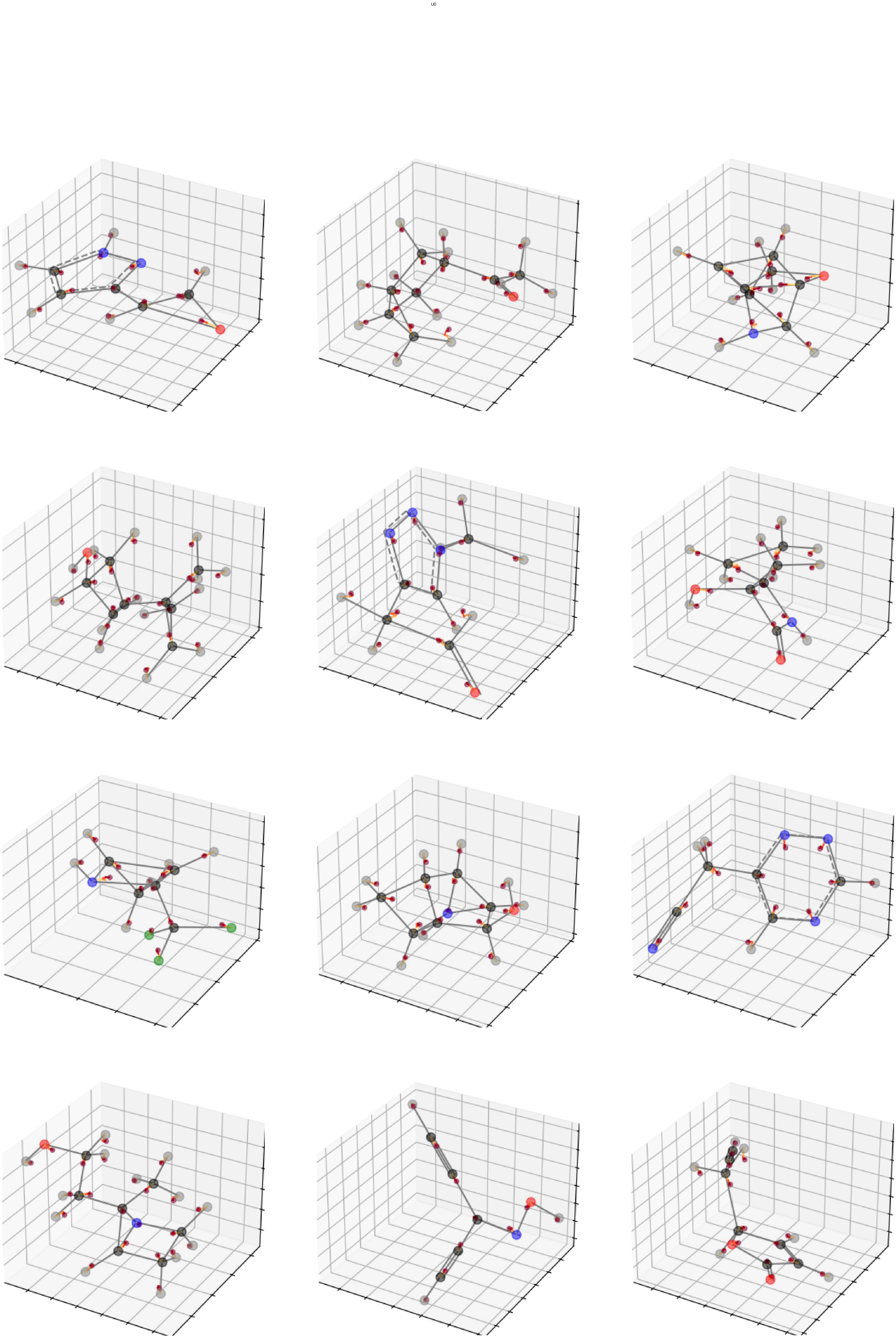}
    \caption{More trajectories of $\tilde{x}$ trained on $U_0$, based on  torchMD-NET.
}
    \label{fig:U0-more-example}
\end{figure}

\newpage

\begin{figure}[hp]
    \centering
    \includegraphics[width=0.92\textwidth]{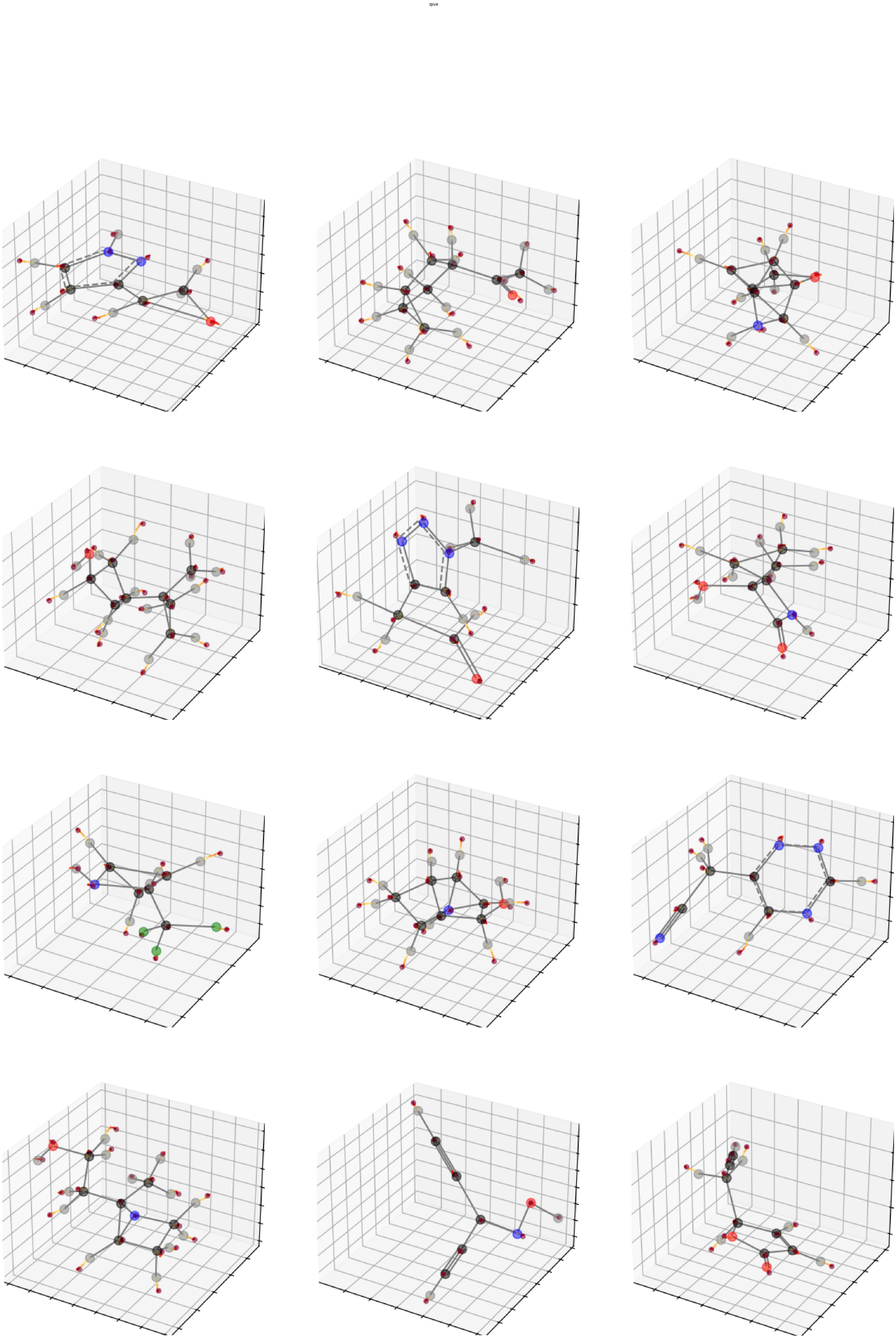}
    \caption{More position of movable position $\tilde{x}$ trained on $zpve$, based on  torchMD-NET.
}
    \label{fig:zpve-more-example}
\end{figure}

\newpage
\subsection{Visualization of atomic and molecular orbitals}
\begin{figure}[h]
    \centering
    \includegraphics[width=0.95\linewidth]{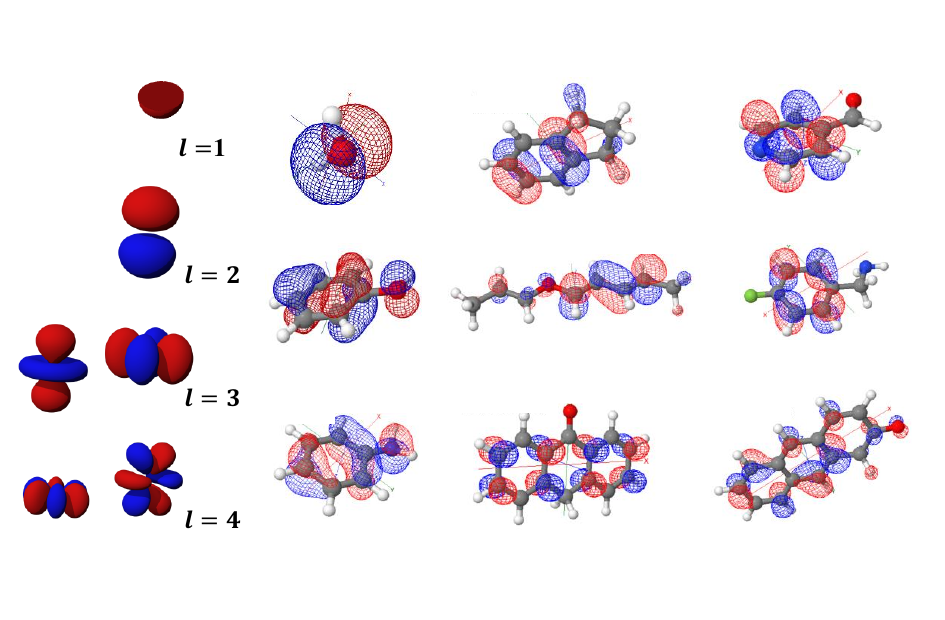}
    \caption{Orbitals of various molecules. The polarization effect is characterized by each molecule type.}
    \label{fig:enter-label}
\end{figure}

%% file: neurips_2024.bbl
\begin{thebibliography}{10}

\bibitem{levine2014quantum}
I.N. Levine.
\newblock Quantum chemistry.
\newblock {\em Pearson advanced chemistry series}, 2014.

\bibitem{bartok2010gaussian}
Albert~P Bart{\'o}k, Mike~C Payne, Risi Kondor, and G{\'a}bor Cs{\'a}nyi.
\newblock Gaussian approximation potentials: The accuracy of quantum mechanics, without the electrons.
\newblock {\em Physical review letters}, 104(13):136403, 2010.

\bibitem{mardirossian2017thirty}
Narbe Mardirossian and Martin Head-Gordon.
\newblock Thirty years of density functional theory in computational chemistry: an overview and extensive assessment of 200 density functionals.
\newblock {\em Molecular Physics}, 115(19):2315--2372, 2017.

\bibitem{yang1991direct}
Weitao Yang.
\newblock Direct calculation of electron density in density-functional theory.
\newblock {\em Physical review letters}, 66(11):1438, 1991.

\bibitem{kryachko2012energy}
Eugene~S Kryachko and Eduardo~V Lude{\~n}a.
\newblock {\em Energy density functional theory of many-electron systems}, volume~4.
\newblock Springer Science \& Business Media, 2012.

\bibitem{andzelm1992density}
J~Andzelm and E~Wimmer.
\newblock Density functional gaussian-type-orbital approach to molecular geometries, vibrations, and reaction energies.
\newblock {\em The Journal of chemical physics}, 96(2):1280--1303, 1992.

\bibitem{mortensen2005real}
Jens~J{\o}rgen Mortensen, Lars~Bruno Hansen, and Karsten~Wedel Jacobsen.
\newblock Real-space grid implementation of the projector augmented wave method.
\newblock {\em Physical review B}, 71(3):035109, 2005.

\bibitem{morris2005real}
Richard~J Morris, Rafael~J Najmanovich, Abdullah Kahraman, and Janet~M Thornton.
\newblock Real spherical harmonic expansion coefficients as 3d shape descriptors for protein binding pocket and ligand comparisons.
\newblock {\em Bioinformatics}, 21(10):2347--2355, 2005.

\bibitem{mchale2017molecular}
Jeanne~L McHale.
\newblock {\em Molecular spectroscopy}.
\newblock CRC Press, 2017.

\bibitem{hariharan1973influence}
Praveen~C Hariharan and John~A Pople.
\newblock The influence of polarization functions on molecular orbital hydrogenation energies.
\newblock {\em Theoretica chimica acta}, 28:213--222, 1973.

\bibitem{ditchfield1971self}
RHWJ Ditchfield, Warren~J Hehre, and John~A Pople.
\newblock Self-consistent molecular-orbital methods. ix. an extended gaussian-type basis for molecular-orbital studies of organic molecules.
\newblock {\em The Journal of Chemical Physics}, 54(2):724--728, 1971.

\bibitem{rassolov20016}
Vitaly~A Rassolov, Mark~A Ratner, John~A Pople, Paul~C Redfern, and Larry~A Curtiss.
\newblock 6-31g* basis set for third-row atoms.
\newblock {\em Journal of Computational Chemistry}, 22(9):976--984, 2001.

\bibitem{helgaker2012recent}
Trygve Helgaker, Sonia Coriani, Poul J{\o}rgensen, Kasper Kristensen, Jeppe Olsen, and Kenneth Ruud.
\newblock Recent advances in wave function-based methods of molecular-property calculations.
\newblock {\em Chemical reviews}, 112(1):543--631, 2012.

\bibitem{karelson1996quantum}
Mati Karelson, Victor~S Lobanov, and Alan~R Katritzky.
\newblock Quantum-chemical descriptors in qsar/qspr studies.
\newblock {\em Chemical reviews}, 96(3):1027--1044, 1996.

\bibitem{jensen2001polarization}
Frank Jensen.
\newblock Polarization consistent basis sets: Principles.
\newblock {\em The Journal of Chemical Physics}, 115(20):9113--9125, 2001.

\bibitem{sanchez1997density}
Daniel S{\'a}nchez-Portal, Pablo Ordejon, Emilio Artacho, and Jose~M Soler.
\newblock Density-functional method for very large systems with lcao basis sets.
\newblock {\em International journal of quantum chemistry}, 65(5):453--461, 1997.

\bibitem{ramakrishnan2014quantum}
Raghunathan Ramakrishnan, Pavlo~O Dral, Matthias Rupp, and O~Anatole von Lilienfeld.
\newblock Quantum chemistry structures and properties of 134 kilo molecules.
\newblock {\em Scientific Data}, 1, 2014.

\bibitem{davidson1986basis}
Ernest~R Davidson and David Feller.
\newblock Basis set selection for molecular calculations.
\newblock {\em Chemical Reviews}, 86(4):681--696, 1986.

\bibitem{liao2022equiformer}
Yi-Lun Liao and Tess Smidt.
\newblock Equiformer: Equivariant graph attention transformer for 3d atomistic graphs.
\newblock In {\em The Eleventh International Conference on Learning Representations}, 2022.

\bibitem{batzner20223}
Simon Batzner, Albert Musaelian, Lixin Sun, Mario Geiger, Jonathan~P Mailoa, Mordechai Kornbluth, Nicola Molinari, Tess~E Smidt, and Boris Kozinsky.
\newblock E (3)-equivariant graph neural networks for data-efficient and accurate interatomic potentials.
\newblock {\em Nature communications}, 13(1):1--11, 2022.

\bibitem{anderson2019cormorant}
Brandon Anderson, Truong-Son Hy, and Risi Kondor.
\newblock Cormorant: Covariant molecular neural networks.
\newblock {\em arXiv preprint arXiv:1906.04015}, 2019.

\bibitem{fuchs2020se}
Fabian~B Fuchs, Daniel~E Worrall, Volker Fischer, and Max Welling.
\newblock Se (3)-transformers: 3d roto-translation equivariant attention networks.
\newblock {\em arXiv preprint arXiv:2006.10503}, 2020.

\bibitem{brandstetter2021geometric}
Johannes Brandstetter, Rob Hesselink, Elise van~der Pol, Erik~J Bekkers, and Max Welling.
\newblock Geometric and physical quantities improve e (3) equivariant message passing.
\newblock In {\em International Conference on Learning Representations}, 2021.

\bibitem{unke2021spookynet}
Oliver~T Unke, Stefan Chmiela, Michael Gastegger, Kristof~T Sch{\"u}tt, Huziel~E Sauceda, and Klaus-Robert M{\"u}ller.
\newblock Spookynet: Learning force fields with electronic degrees of freedom and nonlocal effects.
\newblock {\em Nature communications}, 12(1):7273, 2021.

\bibitem{batatia2022mace}
Ilyes Batatia, David~P Kovacs, Gregor Simm, Christoph Ortner, and G{\'a}bor Cs{\'a}nyi.
\newblock Mace: Higher order equivariant message passing neural networks for fast and accurate force fields.
\newblock {\em Advances in Neural Information Processing Systems}, 35:11423--11436, 2022.

\bibitem{frank2022so3krates}
Thorben Frank, Oliver Unke, and Klaus-Robert M{\"u}ller.
\newblock So3krates: Equivariant attention for interactions on arbitrary length-scales in molecular systems.
\newblock {\em Advances in Neural Information Processing Systems}, 35:29400--29413, 2022.

\bibitem{scott2012group}
William~Raymond Scott.
\newblock {\em Group theory}.
\newblock Courier Corporation, 2012.

\bibitem{thomas2018tensor}
Nathaniel Thomas, Tess Smidt, Steven Kearnes, Lusann Yang, Li~Li, Kai Kohlhoff, and Patrick Riley.
\newblock Tensor field networks: Rotation-and translation-equivariant neural networks for 3d point clouds.
\newblock {\em arXiv preprint arXiv:1802.08219}, 2018.

\bibitem{schutt2021equivariant}
Kristof Sch{\"u}tt, Oliver Unke, and Michael Gastegger.
\newblock Equivariant message passing for the prediction of tensorial properties and molecular spectra.
\newblock In {\em International Conference on Machine Learning}, pages 9377--9388. PMLR, 2021.

\bibitem{gilmer2017neural}
Justin Gilmer, Samuel~S Schoenholz, Patrick~F Riley, Oriol Vinyals, and George~E Dahl.
\newblock Neural message passing for quantum chemistry.
\newblock In {\em International Conference on Machine Learning}, pages 1263--1272. PMLR, 2017.

\bibitem{tholke2022torchmd}
Philipp Th{\"o}lke and Gianni De~Fabritiis.
\newblock Torchmd-net: Equivariant transformers for neural network based molecular potentials.
\newblock {\em arXiv preprint arXiv:2202.02541}, 2022.

\bibitem{kalita2021learning}
Bhupalee Kalita, Li~Li, Ryan~J McCarty, and Kieron Burke.
\newblock Learning to approximate density functionals.
\newblock {\em Accounts of Chemical Research}, 54(4):818--826, 2021.

\bibitem{schleder2019dft}
Gabriel~R Schleder, Antonio~CM Padilha, Carlos~Mera Acosta, Marcio Costa, and Adalberto Fazzio.
\newblock From dft to machine learning: recent approaches to materials science--a review.
\newblock {\em Journal of Physics: Materials}, 2(3):032001, 2019.

\bibitem{engel2011density}
Eberhard Engel.
\newblock {\em Density functional theory}.
\newblock Springer, 2011.

\bibitem{kohn1965self}
Walter Kohn and Lu~Jeu Sham.
\newblock Self-consistent equations including exchange and correlation effects.
\newblock {\em Physical review}, 140(4A):A1133, 1965.

\bibitem{satorras2021n}
V{\i}ctor~Garcia Satorras, Emiel Hoogeboom, and Max Welling.
\newblock E (n) equivariant graph neural networks.
\newblock In {\em International conference on machine learning}, pages 9323--9332. PMLR, 2021.

\bibitem{chmiela2019sgdml}
Stefan Chmiela, Huziel~E Sauceda, Igor Poltavsky, Klaus-Robert M{\"u}ller, and Alexandre Tkatchenko.
\newblock sgdml: Constructing accurate and data efficient molecular force fields using machine learning.
\newblock {\em Computer Physics Communications}, 240:38--45, 2019.

\bibitem{GAMESS}
Giuseppe M.~J. Barca, Colleen Bertoni, Laura Carrington, Dipayan Datta, Nuwan De~Silva, J.~Emiliano Deustua, Dmitri~G. Fedorov, Jeffrey~R. Gour, Anastasia~O. Gunina, Emilie Guidez, Taylor Harville, Stephan Irle, Joe Ivanic, Karol Kowalski, Sarom~S. Leang, Hui Li, Wei Li, Jesse~J. Lutz, Ilias Magoulas, Joani Mato, Vladimir Mironov, Hiroya Nakata, Buu~Q. Pham, Piotr Piecuch, David Poole, Spencer~R. Pruitt, Alistair~P. Rendell, Luke~B. Roskop, Klaus Ruedenberg, Tosaporn Sattasathuchana, Michael~W. Schmidt, Jun Shen, Lyudmila Slipchenko, Masha Sosonkina, Vaibhav Sundriyal, Ananta Tiwari, Jorge~L. Galvez~Vallejo, Bryce Westheimer, Marta Wloch, Peng Xu, Federico Zahariev, and Mark~S. Gordon.
\newblock Recent developments in the general atomic and molecular electronic structure system.
\newblock {\em The Journal of Chemical Physics}, 152(15):154102, April 2020.

\bibitem{schutt2017schnet}
Kristof~T Sch{\"u}tt, Pieter-Jan Kindermans, Huziel~E Sauceda, Stefan Chmiela, Alexandre Tkatchenko, and Klaus-Robert M{\"u}ller.
\newblock Schnet: A continuous-filter convolutional neural network for modeling quantum interactions.
\newblock {\em arXiv preprint arXiv:1706.08566}, 2017.

\bibitem{klicpera_dimenetpp_2020}
Johannes Klicpera, Shankari Giri, Johannes~T. Margraf, and Stephan G{\"u}nnemann.
\newblock Fast and uncertainty-aware directional message passing for non-equilibrium molecules.
\newblock In {\em NeurIPS-W}, 2020.

\bibitem{gasteiger2021gemnet}
Johannes Gasteiger, Florian Becker, and Stephan G{\"u}nnemann.
\newblock Gemnet: Universal directional graph neural networks for molecules.
\newblock {\em Advances in Neural Information Processing Systems}, 34:6790--6802, 2021.

\bibitem{de2018octet}
Johan~J de~Swart.
\newblock The octet model and its clebsch-gordan coefficients.
\newblock In {\em The Eightfold Way}, pages 120--143. CRC Press, 2018.

\bibitem{wigner1931gruppentheorie}
Eugen Wigner.
\newblock Gruppentheorie und ihre anwendung auf die quantenmechanik der atomspektren.
\newblock {\em Monatshefte für Mathematik und Physik}, 1931.

\bibitem{deng2021vector}
Congyue Deng, Or~Litany, Yueqi Duan, Adrien Poulenard, Andrea Tagliasacchi, and Leonidas~J Guibas.
\newblock Vector neurons: A general framework for so (3)-equivariant networks.
\newblock In {\em Proceedings of the IEEE/CVF International Conference on Computer Vision}, pages 12200--12209, 2021.

\bibitem{hoogeboom2022equivariant}
Emiel Hoogeboom, V{\i}ctor~Garcia Satorras, Cl{\'e}ment Vignac, and Max Welling.
\newblock Equivariant diffusion for molecule generation in 3d.
\newblock In {\em International conference on machine learning}, pages 8867--8887. PMLR, 2022.

\bibitem{ryczko2019deep}
Kevin Ryczko, David~A Strubbe, and Isaac Tamblyn.
\newblock Deep learning and density-functional theory.
\newblock {\em Physical Review A}, 100(2):022512, 2019.

\bibitem{schutt2019unifying}
Kristof~T Sch{\"u}tt, Michael Gastegger, Alexandre Tkatchenko, K-R M{\"u}ller, and Reinhard~J Maurer.
\newblock Unifying machine learning and quantum chemistry with a deep neural network for molecular wavefunctions.
\newblock {\em Nature communications}, 10(1):5024, 2019.

\bibitem{pederson2022machine}
Ryan Pederson, Bhupalee Kalita, and Kieron Burke.
\newblock Machine learning and density functional theory.
\newblock {\em Nature Reviews Physics}, 4(6):357--358, 2022.

\bibitem{huang2023central}
Bing Huang, Guido~Falk von Rudorff, and O~Anatole von Lilienfeld.
\newblock The central role of density functional theory in the ai age.
\newblock {\em Science}, 381(6654):170--175, 2023.

\bibitem{xie2018crystal}
Tian Xie and Jeffrey~C Grossman.
\newblock Crystal graph convolutional neural networks for an accurate and interpretable prediction of material properties.
\newblock {\em Physical review letters}, 120(14):145301, 2018.

\bibitem{allam2018application}
Omar Allam, Byung~Woo Cho, Ki~Chul Kim, and Seung~Soon Jang.
\newblock Application of dft-based machine learning for developing molecular electrode materials in li-ion batteries.
\newblock {\em RSC advances}, 8(69):39414--39420, 2018.

\bibitem{lee2022machine}
Andrew Lee, Suchismita Sarker, James~E Saal, Logan Ward, Christopher Borg, Apurva Mehta, and Christopher Wolverton.
\newblock Machine learned synthesizability predictions aided by density functional theory.
\newblock {\em Communications Materials}, 3(1):73, 2022.

\bibitem{chen2022improving}
Pin Chen, Jianwen Chen, Hui Yan, Qing Mo, Zexin Xu, Jinyu Liu, Wenqing Zhang, Yuedong Yang, and Yutong Lu.
\newblock Improving material property prediction by leveraging the large-scale computational database and deep learning.
\newblock {\em The Journal of Physical Chemistry C}, 126(38):16297--16305, 2022.

\bibitem{huang2022provably}
Hsin-Yuan Huang, Richard Kueng, Giacomo Torlai, Victor~V Albert, and John Preskill.
\newblock Provably efficient machine learning for quantum many-body problems.
\newblock {\em Science}, 377(6613):eabk3333, 2022.

\bibitem{huang2023towards}
Bing Huang, Guido~Falk von Rudorff, and O~Anatole von Lilienfeld.
\newblock Towards self-driving laboratories in chemistry and materials sciences: The central role of dft in the era of ai.
\newblock {\em arXiv preprint arXiv:2304.03272}, 2023.

\bibitem{duan2021putting}
Chenru Duan, Fang Liu, Aditya Nandy, and Heather~J Kulik.
\newblock Putting density functional theory to the test in machine-learning-accelerated materials discovery.
\newblock {\em The Journal of Physical Chemistry Letters}, 12(19):4628--4637, 2021.

\bibitem{yu2024qh9}
Haiyang Yu, Meng Liu, Youzhi Luo, Alex Strasser, Xiaofeng Qian, Xiaoning Qian, and Shuiwang Ji.
\newblock Qh9: A quantum hamiltonian prediction benchmark for qm9 molecules.
\newblock {\em Advances in Neural Information Processing Systems}, 36, 2024.

\bibitem{hohenberg1964inhomogeneous}
Pierre Hohenberg and Walter Kohn.
\newblock Inhomogeneous electron gas.
\newblock {\em Physical review}, 136(3B):B864, 1964.

\bibitem{serre1977linear}
Jean-Pierre Serre et~al.
\newblock {\em Linear representations of finite groups}, volume~42.
\newblock Springer, 1977.

\bibitem{inui2012group}
Teturo Inui, Yukito Tanabe, and Yositaka Onodera.
\newblock {\em Group theory and its applications in physics}, volume~78.
\newblock Springer Science \& Business Media, 2012.

\bibitem{bhlitem93362}
Deutsche~Akademie der Wissenschaften~zu Berlin.
\newblock {\em Sitzungsberichte der Königlich Preussischen Akademie der Wissenschaften zu Berlin}, volume Jan-Mai 1882.
\newblock Berlin, Deutsche Akademie der Wissenschaften zu Berlin, 1882-1918, 1882.
\newblock https://www.biodiversitylibrary.org/bibliography/42231.

\bibitem{hall2013lie}
Brian~C Hall and Brian~C Hall.
\newblock {\em Lie groups, Lie algebras, and representations}.
\newblock Springer, 2013.

\bibitem{muller2006spherical}
Claus M{\"u}ller.
\newblock {\em Spherical harmonics}, volume~17.
\newblock Springer, 2006.

\bibitem{shiraishispin}
M~Shiraishi.
\newblock {\em Spin-Weighted Spherical Harmonic Function}.
\newblock Springer, 2013.

\bibitem{chmiela2017machine}
Stefan Chmiela, Alexandre Tkatchenko, Huziel~E Sauceda, Igor Poltavsky, Kristof~T Sch{\"u}tt, and Klaus-Robert M{\"u}ller.
\newblock Machine learning of accurate energy-conserving molecular force fields.
\newblock {\em Science advances}, 3(5):e1603015, 2017.

\bibitem{kipf2018neural}
Thomas Kipf, Ethan Fetaya, Kuan-Chieh Wang, Max Welling, and Richard Zemel.
\newblock Neural relational inference for interacting systems.
\newblock In {\em International conference on machine learning}, pages 2688--2697. PMLR, 2018.

\end{thebibliography}
